\documentclass[12pt,a4]{article}

\usepackage{amssymb}
\usepackage{amsmath}
\usepackage{graphicx}

\begin{document}

\title{Relativistic persistent currents in ideal  Aharonov-Bohm rings and cylinders}

\author{Ion I. Cot\u aescu \\
{\small \it West University of Timi\c{s}oara,}\\
{\small \it V.  P\^{a}rvan Ave.  4, RO-300223 Timi\c{s}oara, Romania}}

\maketitle

\begin{abstract}
The exact solutions of the complete (1+3)-dimensional Dirac equation of fermions moving in  ideal Aharonov-Bohm (AB) rings and cylinders  are used for deriving the exact expressions of the relativistic partial currents. It is shown that  these currents can be related to the derivative of the fermion energy with respect to the flux parameter, just as in the non-relativistic case. However, a new and remarkable relativistic effect is the saturation of the partial currents for high values of the total angular momentum. Based on this property, the total relativistic persistent currents at $T=0$ is evaluated for rings and cylinders obtaining approximative simple closed formulas. 
 
\end{abstract}


\newpage

\section{Introduction}

The electronic effects in mesoscopic rings were studied by using the non-relativistic quantum mechanics \cite{B1}-\cite{B10}  based on the Schr\" odinger  equation with additional terms describing the spin-orbit interaction \cite{B11}-\cite{B15}. 

However, there are  nano-systems, as for example the graphenes,  where several relativistic effects can be observed in the electronic transport. These can be satisfactory explained  considering the electrons as  massless Dirac particles moving on honeycomb lattices \cite{N1}-\cite{Y2}.  Other Dirac materials are the topological insulators like $HgTe$ and $HgTe/CdTe$ quantum wells with low density and high mobility, in which the quantum spin Hall effect can be  realized \cite{Bu1,Bu2,Bu3}. 

Consequently,  many  studies  \cite{N1,N2}, \cite{B16}-\cite{B20} concentrate on the relativistic effects considering the electrons near the Fermi surface as being described by  the $(1+2)$-dimensional Dirac equation corresponding to a restricted three-dimensional Clifford algebra. However, in this manner one restricts simultaneously not only the orbital degrees of freedom but the spin ones too, reducing them to those of the $SO(1,2)$ symmetry.  

Under such circumstances, we believe that there are situations when it is convenient to use the {\em complete} $(1+3)$-dimensional Dirac equation  restricting the orbital motion, according to the concrete geometry of the studied system, but without affecting the natural spin degrees of freedom described by the $SL(2,{\Bbb C})$ group. Thus the polarization effects could be better pointed out. Nevertheless, the complete  Dirac equation was only occasionally used for investigating  some special problems of the fermions in external Aharonov-Bohm (AB) field as for example the spin effects in perturbation theory \cite{scat1,scat2,scat3}, the behaviour of the AB  fermions  in MIT cylinders \cite{MIT} and even  the AB dynamics using numerical methods.  

In the present report we discuss this topics focusing on the results obtained in Refs. \cite{CBC,CBC1}. Herein we have shown first that the solutions of the  Dirac equation in  AB rings and cylinders may be determined as common eigenspinors of a  complete systems of commuting operators including the energy, total angular momentum and a specific operator analogous to the well-known Dirac spherical operator of the relativistic central problems \cite{TH}. These solutions can be normalized with respect to the relativistic scalar product obtaining thus the system of normalized fundamental solutions that allow us to write down the exact expressions of the relativistic partial currents and derive the persistent ones.  The relativistic partial currents we have obtained are related to the derivative of the relativistic energies as in the non-relativistic case but, in contrast with this, there appears a crucial difference, i. e.  in the non-relativistic theory the partial currents are proportional to the angular momentum while in our approach the relativistic currents tend to {\em saturation} in the limit of high total angular momenta. For this reason we have reconsidered the problem of the relativistic persistent currents at $T=0$ proposing an approximative  analytic formula that matches the numerical calculations with a satisfactory accuracy \cite{CBC,CBC1}.

This report  is organized as follows. In the second section we present the relativistic theory of the fermions in AB rings based on a suitable restriction of  the complete Dirac equation, deducing the form of the normalized spinors, deriving the partial currents and writing down the formula of the persistent currents. The next section is devoted to  the relativistic theory of the fermions in AB cylinders based on a restriction of  the complete Dirac equation appropriate to this case. The fundamental solutions on finite or infinite AB cylinders are derived and the partial current are calculated pointing out the saturation effect in both these cases. For the finite cylinders we apply  boundary conditions of MIT type that eliminate the longitudinal current but preserving the saturation properties of the circular one. Finally,  we derive the persistent circular currents at $T=0$ on finite AB cylinders discussing the case of very short ones and the non-relativistic limit.

\section{Dirac fermions in AB rings}

Let us consider a Dirac fermion of mass $M$  moving on a {\em ideal} ring of radius $R$  whose axis is oriented along the homogeneous and static external magnetic field $\vec{B}$ given by the electromagnetic potentials $A_0=0$ and $\vec{A}=\frac{1}{2}\vec{B}\land \vec{x}$ .  

\subsection{The restricted Dirac equation}

The ideal ring is a one-dimensional manifold (without internal structure) embedded in the three-dimensional space according to the equations  $r=R$ and $z=0$,  written in cylindrical coordinates $(t,\vec{x}) \to (t, r, \phi, z)$ with the $z$ axis oriented along $\vec{B}$. Then, it is natural to assume that any field $\psi$ defined on this manifold depends only on the remaining coordinates $(t,\phi)$ such that $\partial_r\psi=0$ and $\partial_z\,\psi=0$. These restrictions give the kinetic term, 
\begin{equation}
{\cal S}_0=\int dt\, d\phi \,\left\{\frac{i}{2}\left[\overline{\psi}(\gamma^0\partial_t\psi +\gamma^{\phi}\partial_{\phi}\psi) -(\partial_t\overline{\psi}\gamma^0+\partial_{\phi}\overline{\psi}\gamma^{\phi})\psi \right]-M\overline{\psi}\psi\right\}\,,
\end{equation}
of the Dirac action ${\cal S}={\cal S}_0-\beta\int dt\, d\phi\, \overline{\psi}\gamma^{\phi}\psi$ in the mentioned external magnetic field, where $\overline{\psi}=\psi^{\dagger}\gamma^0$ and 
\begin{equation}\label{gamph}
\gamma^{\phi}=\frac{1}{R}(-\gamma^1\sin\phi+\gamma^2\cos\phi)
\end{equation}
is depending on $\phi$. The notation $\beta=\frac{1}{2}eBR^2$ stands for the usual dimensionless flux parameter (in natural units).

From this action we obtain  the correctly restricted Dirac equation, $E_D\psi=M\psi$,  with  the  new self-adjoint Dirac operator   
\begin{equation}\label{D2}
E_D=i\gamma^0\partial_t +\gamma^{\phi}(i\partial_{\phi}-\beta)+\frac{i}{2}\,\partial_{\phi}(\gamma^{\phi})\,,
\end{equation}
whose supplemental last term guarantees that $\overline{E}_D=E_D$. 
This operator commutes with the energy operator $H=i\partial_t$ and the third component, $J_3=L_3+S_3$, of the total angular momentum, formed by the orbital part $L_3=-i\partial_{\phi}$ the spin one $S_3=\frac{1}{2}\,{\rm diag} (\sigma_3,\sigma_3)$. Therefore, we have the opportunity to look for particular solutions of the form 
\begin{equation}\label{psi}
\psi_{E,\lambda}(t, \phi)=N\left(
\begin{array}{c}
f_1 e^{i\phi(\lambda-\frac{1}{2})}\\
f_2 e^{i\phi(\lambda+\frac{1}{2})}\\
g_1 e^{i\phi(\lambda-\frac{1}{2})}\\
g_2 e^{i\phi(\lambda+\frac{1}{2})}
\end{array}\right) e^{-iEt}\,,
\end{equation}
which satisfy the common eigenvalue problems, $E_D\psi_{E,\lambda}(t, \phi)=M\psi_{E,\lambda}(t, \phi)$ and
\begin{equation}
H\psi_{E,\lambda}(t, \phi)=E\psi_{E,\lambda}(t, \phi)\,, \quad  
J_3\psi_{E,\lambda}(t, \phi)=\lambda\psi_{E,\lambda}(t, \phi)\,,
\end{equation}
laying out the energy $E$ and  the angular quantum number $\lambda=\pm\frac{1}{2}\pm \frac{3}{2},...$ whose values are determined by the condition $\psi_{E,\lambda}(t, \phi+2\pi)=\psi_{E,\lambda}(t, \phi)$.

In this manner we separated the variables remaining with a system of algebraic equations  that  in the standard representation of the gamma matrices (with diagonal $\gamma^0$) reads
\begin{equation}
\left(
\begin{array}{cccc}
E-M &0&0 &\frac{i}{R}(\lambda+\beta)\\
0&E-M &-\frac{i}{R}(\lambda+\beta)&0\\
0&-\frac{i}{R}(\lambda+\beta)&-E-M&0\\
\frac{i}{R}(\lambda+\beta)&0&0&-E-M
\end{array}\right)\,\left(
\begin{array}{c}
f_1\\
f_2\\
g_1\\
g_2
\end{array}\right) =0\,.
\end{equation}
This system has non-trivial solutions only for the discrete values of energy  
\begin{equation}\label{ene}
E_{\lambda}=
\frac{1}{R}\left[M^2R^2+(\beta+\lambda)^2\right]^{\frac{1}{2}}\,,
\end{equation}
whose second terms encapsulate the AB effect. For each value $E_{\lambda}$ we find  two particular solutions  for which
\begin{equation}
\left(
\begin{array}{c}
f_1\\
f_2
\end{array}\right)= \xi_{\sigma} \,,
\end{equation} 
where $\xi_{\sigma}$ are the usual Pauli spinors of polarization $\sigma=\pm\frac{1}{2}$ with respect to the $z$ axis,
\begin{equation}
\xi_{\frac{1}{2}}=\left(
\begin{array}{c}
1\\
0
\end{array}\right)\,,\quad \xi_{-\frac{1}{2}}=\left(
\begin{array}{c}
0\\
1
\end{array}\right)\,.
\end{equation}
Thus,  we find that for $\sigma=\frac{1}{2}$ the spinors (\ref{psi})
take the form
\begin{equation}\label{psiUp0}
U_{\lambda}^+(t, \phi)=\frac{1}{2\sqrt{\pi E_{\lambda}R}}\left(
\begin{array}{c}
\sqrt{E_{\lambda}-M}\,    e^{i\phi(\lambda-\frac{1}{2})}\\
0\\
0\\
i\sqrt{E_{\lambda}+M}\,e^{i\phi(\lambda+\frac{1}{2})}
\end{array}\right) e^{-iE_{\lambda}t}\,,
\end{equation} 
while for $\sigma=-\frac{1}{2}$ we obtain the solutions
\begin{equation}\label{psiUm0}
U_{\lambda}^-(t,\phi)=\frac{1}{2\sqrt{\pi E_{\lambda}R}}\left(
\begin{array}{c}
0\\
\sqrt{E_{\lambda}-M}\,e^{i\phi(\lambda+\frac{1}{2})}\\
-i \sqrt{E_{\lambda}+M}\,e^{i\phi(\lambda-\frac{1}{2})}\\
0
\end{array}\right) e^{-iE_{\lambda}t}\,.
\end{equation}  
The normalization constants are fixed in accordance to the  relativistic scalar product
\begin{equation}
\langle \psi, \psi'\rangle=R\,\int_{0}^{2\pi}d\phi\,\psi^{\dagger}(t,\phi) \psi'(t,\phi)\,,
\end{equation} 
such that 
\begin{equation}
\langle U^{\pm}_{\lambda}, U^{\pm}_{\lambda'}\rangle=\delta_{\lambda,\lambda'}\,,\quad \langle U^{\pm}_{\lambda}, U^{\mp}_{\lambda'}\rangle=0\,.
\end{equation}

Hence we obtained a pair of fundamental solutions of the same energy and total angular momentum but which are not eigenspinors of the operators $L_3$ or $S_3$. Therefore, we may ask how these solutions can be defined as different eigenspinors of a new operator. The answer is obvious if we observe that the desired operator is  $K=2\gamma^0 S_3$ which satisfies
$K U^{\pm}_{\lambda}=\pm U^{\pm}_{\lambda}$. The conclusion is that the spinors  $U^{\pm}_{\lambda}$ are  common eigenspinors of the complete set of commuting operators $\{E_D,H,K,J_3\}$.
    
The operator $K$ introduced above is the analogous of the spherical Dirac operator $K_D=\gamma^0(2\vec{S}\cdot\vec{L}+1)$ that concentrates the angular variables of the Dirac equation in external fields with central symmetry \cite{TH}. Note that the genuine three-dimensional operator $K_D$ cannot be used here because of our dimensional reduction such that we must consider the simplified version $K$ \footnote{It is known that the forms of such operators depend on the number of space dimensions \cite{Dong}}. The eigenvalues of this operator give the polarization in the non-relativistic limit. For this reason we keep this terminology considering that the eigenvalues $\kappa=\pm1$ of the operator $K$ define the fermion polarization with respect to the direction of the magnetic field $\vec{B}$.

\subsection{Relativistic currents in AB rings}

Using the above results we can calculate the exact relativistic expressions of the partial  currents on quantum rings, pointing out the difference between the genuine relativistic theory and  the non-relativistic one. We show that in the relativistic approach the partial current tends to saturation for increasing $\lambda$ such that the persistent currents at $T=0$ will get new properties.

Let us start with the quantum rings where the states of the fermions of energy $E_\lambda$ are  described by the normalized linear combinations
\begin{equation}
\psi_{\lambda}=c_+U^+_{\lambda}+c_-U^-_{\lambda}\,, \quad |c_+|^2+|c_-|^2=1\,,
\end{equation} 
for which  the expectation value of the polarization operator reads,
\begin{equation}
\langle \psi_{\lambda}, K\psi_{\lambda}\rangle =|c_+|^2-|c_-|^2\,.
\end{equation}

The partial  currents (of given $\lambda$) coincide in this case with their  densities,     
$I_{\lambda}=R\,\overline{\psi}_{\lambda}\gamma^{\phi}\psi_{\lambda}$,
that can be calculated with the help of the matrix (\ref{gamph}). Then, observing that 
\begin{equation}\label{inter}
{\overline{U}_{\lambda}^{\pm}}(t,\phi)\gamma^{\phi}U_{\lambda}^{\mp}(t,\phi)=0\,,
\end{equation}
we obtain the {\em partial}  current of a fermion of energy $E_{\lambda}$ as
\begin{equation}
I_{\lambda}=|c_+|^2 I^+_{\lambda}+|c_-|^2I_{\lambda}^-=\frac{1}{2\pi R^2}\frac{\beta+\lambda}{E_{\lambda}}=\frac{1}{2\pi}\frac{\partial E_{\lambda}}{\partial\beta}\,,
\end{equation}
since  $I^{\pm}_{\lambda}=R\,{\overline{U}_{\lambda}^{\pm}}(t,\phi)\gamma^{\phi}U_{\lambda}^{\pm}(t,\phi)=I_{\lambda}$. Thus we find that the partial currents are independent on polarization being related to energies in a similar manner as in the non-relativistic theory.

The  exact relativistic expressions of the partial  currents  we obtained here depend only on two dimensionless parameters $\nu=\beta+\lambda$ and  $\mu=MR$ (or $MRc/\hbar$ in usual units) that are the arguments of the auxiliary function $\chi$ defined as
\begin{equation}\label{Irel}
I_{\lambda}=\frac{1}{2\pi R}\, \chi (\mu,\nu)\,, \quad \chi(\mu,\nu)=\frac{\nu}{\sqrt{\mu^2+\nu^2}}\,.
\end{equation} 
This function  has the remarkable  asymptotic behaviour 
\begin{equation}
\lim_{\nu\to\pm\infty}\chi(\mu,\nu)=\pm 1\,,
\end{equation}
which shows that the relativistic partial  currents tend to {\em saturation} for large values of $\lambda$.  Moreover, for small values of $\nu$ we can expand
\begin{equation}\label{cucu}
\chi(\mu,\nu)= \frac{\nu}{\mu}+O(\nu^3)\,.
\end{equation}
Note that the non-relativistic limit recovers the well-known behaviours
\begin{equation}
E_{\lambda}-M \to \tilde E_{\lambda}=\frac{\nu^2}{2 R\mu}\,, \quad I_{\lambda}\to \tilde I_{\lambda}=\frac{1}{2\pi R}\frac{\nu}{\mu}=\frac{1}{2\pi}\frac{\partial \tilde E_{\lambda}}{\partial \nu}\,.
\end{equation} 

Hereby we conclude  that the principal difference is that the relativistic partial currents (\ref{Irel}) are saturated  while in the non-relativistic case we do not meet this effect since the function $\chi(\mu,\nu)$ is replaced then by the  linear function $\frac{\nu}{\mu}$  that  is just its  tangent in $\nu=0$  as we deduce from Eq. (\ref{cucu}). This result was previously outlined in Ref. \cite{Ghosh} but based on the non-Hermitian Dirac equation of Ref. \cite{CP}.
Obviously, the correct saturation effect is given by the expression of the partial currents  (\ref{Irel}) derived here.  
 
Note that the non-relativistic approximation can be used with a satisfactory accuracy  only in the domain where the function $\chi(\mu,\nu)$ is  approaching to the linear function $\frac{\nu}{\mu}$. Our numerical evaluations show that  in the domain  $-\frac{1}{2}\mu<\nu<\frac{1}{2}\mu$  the difference $|\chi(\mu,\nu)-\frac{\nu}{\mu}|$ is satisfactory small remaining less than  0.05.   In addition,  we estimate that for $|\nu|\ge 5\mu$ the current is approaching to its saturation value since $ |\chi(\mu,\pm 5\mu)|= 0.98058$.

\subsection{Relativistic persistent currents in AB rings}

The above results allow us to derive the total persistent current at $T=0$ in a semiconductor ring of parameter $\mu$ having a even number of electrons $N_e$ fixed by the Fermi-Dirac statistics. For the mesoscopic rings with $R=100 {\rm nm}$ the parameter $\mu$ is of the order $10^3 - 10^5$. For example, in a  $InSb$ ring of this radius,  the effective electron mass is $M=m^*_e=0.0135\, m_e$  \cite{Vag} such that  $\mu=3495$. This  seems to be the minimal value of $\mu$  obtained so far but  it is possible to obtain smaller values  in further experiments with mesoscopic rings with $R< 100 {\rm nm}$ or even with nano-rings having $R\sim 10 {\rm nm}$.   According to our estimation, the relativistic effects may be measurable for $\mu<10^3$ which means that the actual experiments  are approaching to this threshold which could be reached soon.

In all these cases the flux parameter $\beta$ remains very small (less than $10^{-8}$) such that we can neglect the terms of the order $O(\beta^2)$ of the Taylor expansions of our functions that depend on  $\nu=\lambda+\beta$.  The total persistent current at $T=0$  is given by the sum  
\begin{equation}
I= \sum_{\lambda=-\lambda_F}^{\lambda_F}I_{\lambda}= \sum_{\lambda=\frac{1}{2}}^{\lambda_F}\left(I_{\lambda}+I_{-\lambda}\right)=\frac{1}{2\pi R}\sum_{\lambda=\frac{1}{2}}^{\lambda_F}[\chi(\mu,\lambda+\beta)+\chi(\mu,-\lambda+\beta)]
\end{equation}
over all the allowed polarizations, $\lambda=\pm\frac{1}{2},\pm\frac{3}{2},...,\pm\lambda_F$ where $\lambda_F=\frac{1}{2}(N_e-1)$.  Furthermore,  by using the expansion
\begin{equation}
\chi(\mu,\lambda+\beta)+\chi(\mu,-\lambda+\beta)=2 j(\mu,\lambda) \beta +O(\beta^3)\,,
\end{equation}
where
\begin{equation}
j(\mu,\lambda)=\frac{\mu^2}{(\mu^2+\lambda^2)^{\frac{3}{2}}}\,,
\end{equation}
we arrive at the relativistic persistent currents,  
\begin{equation}\label{Ic}
I=c(\mu) I_{max}\,,\quad I_{max}=\frac{\beta}{\pi R}\,, \quad c(\mu)=\sum_{\lambda=\frac{1}{2}}^{\lambda_F}j(\mu,\lambda)\,,
\end{equation}
that can be calculated numerically on computer for any concrete value of $\mu$.

{ \begin{figure}
    \centering
    \includegraphics[scale=0.37]{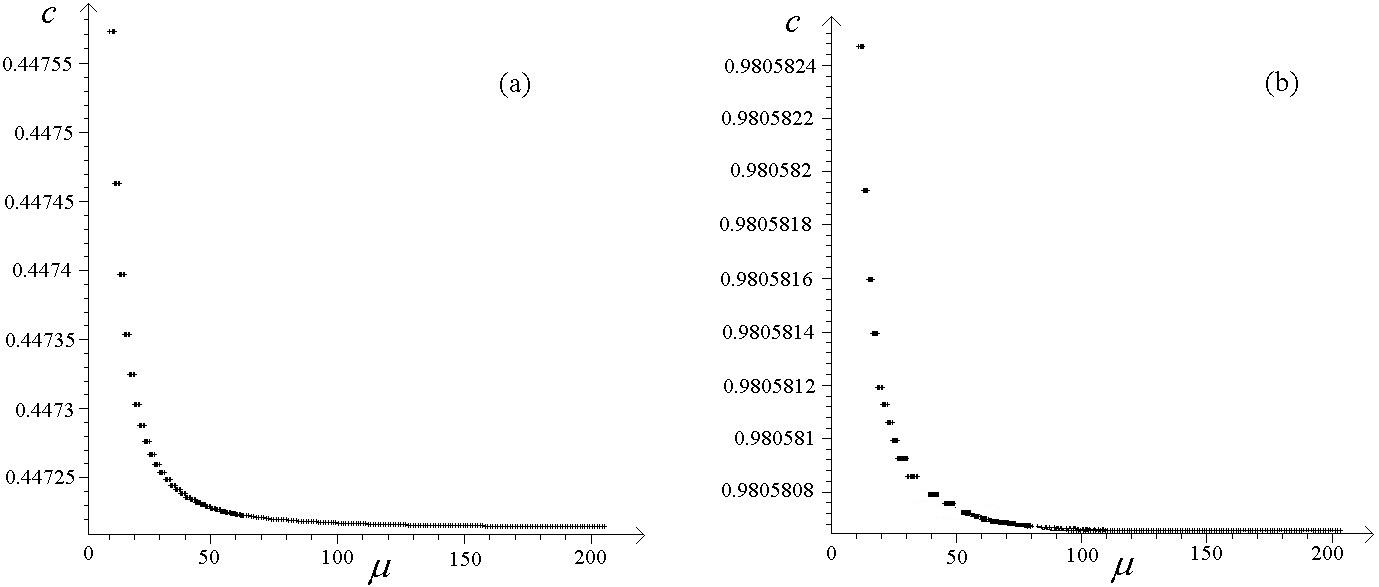}
    \caption{The function $c(\mu)$ versus $\mu$ calculated for $\lambda_F=0.5\mu$ (a) and $\lambda_F=5 \mu$  when the co-domain is very narrow, $\sim 10^{-5}$  (b).  }
  \end{figure}}

The function $j(\mu,\lambda)$ is simple reaching its maximal value 0.7698 for $\mu=\frac{1}{\sqrt{2}}$ and $\lambda=\frac{1}{2}$ and  decreasing  then monotonously  to zero when $\mu$ and $\lambda$ are increasing to infinity.  This behaviour is a direct consequence of the saturation of the partial currents that compensate each other in the saturation zone where $I_{\lambda}+I_{-\lambda} \to 0$.   These simple monotony and smoothness properties of the function $j(\mu,\lambda)$ lead to nice results concerning the values of the sum (\ref{Ic}c) when we compute all the allowed contributions. Our numerical examples show that when $\mu$ is increasing then the functions $c(\mu)$ are monotonously decreasing tending to an asymptotic value (as in Fig. 1). Consequently, in the asymptotic zone, $\mu>100$, we can use the following approximation
\begin{equation}\label{Intc}
c(\mu)\simeq \int_{0}^{\lambda_F}j(\mu,\lambda)d\lambda= \frac{\lambda_F}{\sqrt{\mu^2+\lambda_F^2}}\,,
\end{equation}
giving the definitive formula of the relativistic persistent currents
\begin{equation}\label{final}
I=\frac{k}{\sqrt{1+k^2}}\, I_{max}\,, \quad k=\frac{\lambda_F}{\mu}\simeq \frac{N_e}{2\mu}\,,
\end{equation} 
that reproduces the numerical results with a satisfactory accuracy (under $10^{-5}$). 
Note that the non-relativistic persistent current, that in our notation reads $\tilde I=k I_{max}$, represents a good approximation of Eq. (\ref{final}) only for small values of $k$ (say $k<0.2$) for which we can use the approximation $k(1+k^2)^{-\frac{1}{2}}=k+ O(k^3)\simeq k$. 

\section{Dirac fermions on AB cylinders}

We consider the motion of a Dirac fermion of mass $M$  on an {\em ideal} cylinder of radius $R$  whose axis is oriented along the homogeneous and static external magnetic field $\vec{B}$ given by the electromagnetic potentials $A_0=0$ and $\vec{A}=\frac{1}{2}\vec{B}\land \vec{x}$.  This background  is a two-dimensional manifold (without internal structure) embedded in the three-dimensional space obeying the simple equation  $r=R$  in cylindrical coordinates $(t,\vec{x}) \to (t, r, \phi, z)$ with the $z$ axis oriented along $\vec{B}$. 

\subsection{The restricted Dirac equation}

Then, it is natural to assume that any field $\psi$ defined on this manifold depends only on the remaining coordinates $(t,\phi,z)$ such that we can put  $\partial_r\psi=0$ in the kinetic term of the Lagrangian density. Thus we obtain the action of the Dirac fermion in the mentioned external magnetic field
\begin{equation}
{\cal S}={\cal S}_0-\beta\int dt\, d\phi\, dz\,\overline{\psi}\gamma^{\phi}\psi
\end{equation}
having the kinetic part  
\begin{eqnarray}
{\cal S}_0&=&\int dt\, d\phi \,dz\,\left\{\frac{i}{2}\left[\overline{\psi}(\gamma^0\partial_t\psi +\gamma^{\phi}\partial_{\phi}\psi+\gamma^3\partial_z\psi)\right.\right.\nonumber\\ &&-\left.\left.(\partial_t\overline{\psi}\gamma^0+\partial_{\phi}\overline{\psi}\gamma^{\phi}+\partial_z\overline{\psi}\gamma^3)\psi \right]-M\overline{\psi}\psi\right\}\,,
\end{eqnarray}
where  $\overline{\psi}=\psi^{\dagger}\gamma^0$ and 
\begin{equation}\label{gamph}
\gamma^{\phi}=\frac{1}{R}(-\gamma^1\sin\phi+\gamma^2\cos\phi)\,.
\end{equation}
The notation $\beta=\frac{1}{2}eBR^2$ stands for the usual dimensionless flux parameter (in natural units).

From this action we obtain  the correctly restricted Dirac equation, $E_D\psi=M\psi$,  with  the  self-adjoint Dirac operator   
\begin{equation}\label{D2}
E_D=i\gamma^0\partial_t +\gamma^{\phi}(i\partial_{\phi}-\beta)+\frac{i}{2}\,\partial_{\phi}(\gamma^{\phi})+i\gamma^3\partial_z\,,
\end{equation}
whose supplemental third term guarantees that $\overline{E}_D=E_D$. 
This operator commutes with the energy operator $H=i\partial_t$, the momentum along to the third axis, $P_3=-i\partial_z$, and the similar component, $J_3=L_3+S_3$, of the total angular momentum, formed by the orbital part $L_3=-i\partial_{\phi}$ the spin one $S_3=\frac{1}{2}\,{\rm diag} (\sigma_3,\sigma_3)$. 

 Under such circumstances, we have the opportunity to look for particular solutions of the form
\begin{equation}\label{psi}
\psi_{E,\lambda}(t, \phi,z)=N\left(
\begin{array}{c}
f_1(z)e^{i\phi(\lambda-\frac{1}{2})}\\
f_2(z)e^{i\phi(\lambda+\frac{1}{2})}\\
g_1(z)e^{i\phi(\lambda-\frac{1}{2})}\\
g_2(z)e^{i\phi(\lambda+\frac{1}{2})}
\end{array}\right) e^{-iEt}
\end{equation}
which satisfy the eigenvalue problems,
\begin{equation}
H\psi_{E,\lambda}(t, \phi,z)=E\psi_{E,\lambda}(t, \phi,z)\,, \quad
J_3\psi_{E,\lambda}(t, \phi,z)=\lambda\psi_{E,\lambda}(t, \phi,z)\,,
\end{equation}
laying out the energy $E$ and  the quantum number $\lambda=\pm\frac{1}{2},\pm \frac{3}{2},...$ of the total angular momentum.
The normalization constant $N$ has to be determined after we solve the functions of $z$ from the remaining reduced equation that  in the standard representation of the gamma matrices (with diagonal $\gamma^0$) reads
\begin{equation}
\left(
\begin{array}{cccc}
E-M &0&i\partial_z &\frac{i}{R}(\lambda+\beta)\\
0&E-M &-\frac{i}{R}(\lambda+\beta)&-i\partial_z\\
-i\partial_z&-\frac{i}{R}(\lambda+\beta)&-E-M&0\\
\frac{i}{R}(\lambda+\beta)&i\partial_z&0&-E-M
\end{array}\right)\,\left(
\begin{array}{c}
f_1(z)\\
f_2(z)\\
g_1(z)\\
g_2(z)
\end{array}\right) =0\,.
\end{equation}
This is in fact a system of linear differential equations allowing us to solve the functions of $z$.

The general solutions of this system can be obtained reducing the number of functions with the help of the last two equations that yield
\begin{equation}\label{g12}
\left(
\begin{array}{c}
g_1(z)\\
g_2(z)
\end{array}\right)=\frac{1}{E+M}\left(
\begin{array}{cc}
-i\partial_z& -\frac{i}{R}(\lambda+\beta)\\
\frac{i}{R}(\lambda+\beta)&i\partial_z
\end{array}\right)\left(
\begin{array}{c}
f_1(z)\\
f_2(z)
\end{array}\right)
\end{equation}
leading to the second order equations
\begin{equation}\label{ene}
\left(E^2-M^2-\frac{1}{R^2}(\lambda+\beta)^2 +\partial_z^2\right)f_{1,2}(z)=0\,.
\end{equation}
Consequently, the solutions must be linear combinations of the form
\begin{equation}
f_{1,2}(z)=c_{1,2} e^{ikz}+c'_{1,2}e^{-ikz}
\end{equation}
 where $k$ is the fermion momentum along the $z$ axis.  The concrete form of these solutions depends on the boundary conditions we chose for determining  the integration constants $c_{1,2}$ and $c'_{1,2}$ up to a normalization factor, $N$.  This last constant has to be determined by imposing the desired  normalization condition with respect to the relativistic scalar product
 \begin{equation}
 \langle \psi, \psi'\rangle=R\int_{0}^{2\pi}d\phi\int_{D_z}dz \psi^{\dagger}(t,\phi,z) \psi'(t,\phi,z)\,,
 \end{equation}
calculated on the domain $D_z$ of the entire cylinder.

\subsection{Currents on infinite AB cylinders}

The simplest case is of the infinite cylinder, with $D_z={\Bbb R}$,  where the motion along its axis is a free one. We assume that the spin projections are measured just with respect to this axis such that  we may chose
\begin{equation}
\left(
\begin{array}{c}
f_1(z)\\
f_2(z)
\end{array}\right)=Ne^{ikz} \xi_{\sigma}\,, \quad k\in {\Bbb R}\,,
\end{equation}
where $\xi_{\sigma}$ are the usual Pauli spinors, 
\begin{equation}
\xi_{\frac{1}{2}}=\left(
\begin{array}{c}
1\\
0
\end{array}\right)\,,\quad \xi_{-\frac{1}{2}}=\left(
\begin{array}{c}
0\\
1
\end{array}\right)\,.
\end{equation}
 of polarizations $\sigma=\pm\frac{1}{2}$. Then the functions $g_{1,2}(z)$ can be derived from Eq. (\ref{g12}) as
\begin{equation}
\left(
\begin{array}{c}
g_1(z)\\
g_2(z)
\end{array}\right)=\frac{ Ne^{ikz}}{E+M}\left(
\begin{array}{cc}
 k& -\frac{i}{R}(\lambda+\beta)\\
\frac{i}{R}(\lambda+\beta)&- k
\end{array}\right) \xi_{\sigma}\,,
\end{equation}
while the energy that depends on $k$ and $\lambda$ reads
\begin{equation}\label{Ekm}
E_{k,\lambda}=\left[ M^2+k^2+\frac{1}{R^2}\left(\lambda+\beta \right)^2  \right]^{\frac{1}{2}} 
\end{equation}
as it results from Eq. (\ref{ene}). We obtain thus a mixed energy spectrum whose ground level is given by $k=0$ and one of the values $\lambda=\pm\frac{1}{2}$ that minimizes the last term in Eq. (\ref{Ekm}), e. g. $\lambda=-\frac{1}{2}$ if $\beta>0$. Consequently, the spinor components have to be tempered distributions that may be normalized in the momentum scale.

According to the above results we can write  two types of fundamental solutions   (\ref{psi}) of the form 
\begin{equation}
U_{k,\lambda}^{\pm}(t,\phi,z)=u^{\pm}_{k,\lambda}(\phi)\frac{1}{\sqrt{2\pi}}e^{-iE_{k,\lambda}t +ikz}\,,
\end{equation}
corresponding to the polarizations $\sigma =\pm\frac{1}{2}$. For $\sigma =\frac{1}{2}$ we obtain
\begin{equation}\label{psiUp}
u_{k,\lambda}^+(\phi)=N_{k,\lambda}^+\left(
\begin{array}{c}
e^{i\phi(\lambda-\frac{1}{2})}\\
0\\
\frac{k}{E_{k,\lambda}+M}e^{i\phi(\lambda-\frac{1}{2})}\\
\frac{i(\lambda+\beta)}{R(E_{k,\lambda}+M)}e^{i\phi(\lambda+\frac{1}{2})}
\end{array}\right) 
\end{equation}
assuming that the normalization factor depend on $k$ and $\lambda$.
Similarly, for $\sigma=-\frac{1}{2}$ we deduce
\begin{equation}\label{psiUm}
u_{k,\lambda}^-( \phi)=N_{k,\lambda}^-\left(
\begin{array}{c}
0\\
e^{i\phi(\lambda+\frac{1}{2})}\\
\frac{-i(\lambda+\beta)}{R(E_{k,\lambda}+M)}e^{i\phi(\lambda-\frac{1}{2})}\\
\frac{-k}{E_{k,\lambda}+M}e^{i\phi(\lambda+\frac{1}{2})}
\end{array}\right) 
\end{equation}
It remains to calculate the normalization in the momentum scale finding that by fixing the values
\begin{equation}
N_{k,\lambda}^{\pm}=\frac{1}{\sqrt{2\pi R}}\sqrt{\frac{E_{k,\lambda}+M}{2E_{k,\lambda}}}\,,
\end{equation}
we obtain the desired (generalized) orthogonality relations
\begin{equation}
\langle U^{\pm}_{k,\lambda}, U^{\pm}_{k',\lambda'}\rangle=\delta_{\lambda,\lambda'}\delta(k-k')\,,\quad \langle U^{\pm}_{k,\lambda}, U^{\mp}_{k',\lambda'}\rangle=0\,.
\end{equation}

These fundamental solutions are  eigenspinors of the same set of commuting operators which seems to be incomplete as long as we cannot distinguish between $U^+$ and $U^-$. Therefore we need to introduce a new operator for completing this set. A short inspection suggests that this must be an analogous of the Dirac spheric operator \cite{TH} that reads now  $K=\gamma^0(2 S_3 L_3+\frac{1}{2})$ giving the eigenvalues problems  $K U_{k,\lambda}^{\pm} =\pm\lambda U_{k,\lambda}^{\pm}$. The conclusion is that the fundamental solutions we derived above are eigenspinors of the complete set of commuting operators $\{H,K,J_3, P_3\}$.

The  spinors describing physical states are square integrable packets of a given total angular momentum $\lambda$ having the form
\begin{equation}\label{psilam}
\psi_{\lambda}=\int_{-\infty}^{\infty}dk \left[a_+(k) U^+_{k,\lambda}+a_-(k) U^-_{k,\lambda}\right]
\end{equation} 
where the functions $a_{\pm}$ satisfy the  condition
\begin{equation}\label{Calpha}
\int_{-\infty}^{\infty}dk \left[|a_+(k)|^2+|a_-(k)|^2\right]=1
\end{equation}
that assures the normalization condition $\langle \psi_{\lambda}, \psi_{\lambda}\rangle=1$. We say that the packets (\ref{psilam}) describe the states $(\lambda,a)$ in which the expectation value  of the total angular momentum reads $\langle \psi_{\lambda}, J_3 \psi_{\lambda}\rangle= \lambda$ while the polarization degree can be defined as
\begin{equation}
{\cal P}=\langle \psi_{\lambda}, K\psi_{\lambda}\rangle=\lambda\int_{-\infty}^{\infty}dk \left[|a_+(k)|^2-|a_-(k)|^2\right]\,.
\end{equation}

Now we can derive the currents of the fermions in the states $(\lambda,a)$  by using the components of the current density $j_{\lambda}^{\mu}=\overline{\psi}_{\lambda}\gamma^{\mu}\psi_{\lambda}$.  We consider first the total circular current 
\begin{equation}
I^c_{\lambda}=R\int_{-\infty}^{\infty} dz \overline{\psi}_{\lambda}\gamma^{\phi}\psi_{\lambda}\,,
\end{equation} 
that can be calculated according to Eqs. (\ref{gamph}), (\ref{psiUp}) and (\ref{psiUm}), that yield 
\begin{equation}
\overline{u}^{\pm}_{\lambda}\gamma^{\phi}{u}^{\pm}_{\lambda}=\frac{\lambda+\beta}{2\pi R^3 E_{k,\lambda}}\,, \quad \overline{u}^{\pm}_{\lambda}\gamma^{\phi}{u}^{\mp}_{\lambda}=0\,.
\end{equation}
Then, observing that the integral over the $z$ axis generates a $\delta$-function, we obtain  the definitive closed form
\begin{equation}
I_{\lambda}^c=\frac{\lambda+\beta}{2\pi R^2}\int_{-\infty}^{\infty}\frac{dk}{E_{k,\lambda}} \left[|a_+(k)|^2+|a_-(k)|^2\right]\,.
\end{equation}
Hereby we draw the conclusion  that the circular current is stationary (i. e. independent on $t$)
depending only on the packet content as given by the arbitrary functions $a_{\pm}$. 

Nevertheless, it is remarkable that there are two important properties of the circular currents that are independent on the form of  these functions. The first one is the saturation effect for increasing total angular momenta,
\begin{equation}
\lim_{\lambda \to \pm \infty}I^c_{\lambda}=\pm \frac{1}{2\pi R}\int_{-\infty}^{\infty}dk \left[|a_+(k)|^2+|a_-(k)|^2\right] =\pm \frac{1}{2\pi R}\,.
\end{equation}  
On the other hand,  bearing in mind that the expectation value of the energy in the state $(\lambda,a)$  reads
\begin{equation}
E_{\lambda}=\int_{-\infty}^{\infty}{dk}{E_{k,\lambda}} \left[|a_+(k)|^2+|a_-(k)|^2\right]\,,
\end{equation}
we recover the familiar formula
\begin{equation}
I^c_{\lambda}=\frac{1}{2\pi}\frac{\partial E_{\lambda}}{\partial\beta}\,,
\end{equation}
that has the same form as in the case of the relativistic  \cite{CBC} or non-relativistic AB rings.

Note that, in contrast to the circular current, the longitudinal one,
\begin{equation}
I^3_{\lambda}=R\int_{0}^{2\pi} d\phi\, \overline{\psi}_{\lambda}\gamma^{3}\psi_{\lambda}\,,
\end{equation} 
depends on time, reflecting thus the propagation and dispersion of the packet along the $z$ axis. The general expression of this current  is presented in the Appendix A.

\subsection{Currents on finite AB cylinders}

Another interesting problem is of a finite cylinder of length $L$ for which we must consider the boundary conditions $f_{1,2}(0)=f_{1,2}(L)=0$. Therefore, we may chose
\begin{equation}\label{fkn1}
\left(
\begin{array}{c}
f_1(z)\\
f_2(z)
\end{array}\right)= N \sin( k_nz)\,\xi_{\sigma} \,,\quad k_n=\frac{\pi n}{L}\,,n=1,2,...\,,
\end{equation}
denoting now $E_{n,\lambda}=E_{k_n,\lambda}$. Thus we obtain the countable discrete energy spectrum 
\begin{equation}\label{Enm}
E_{n,\lambda}=\left[ M^2+\frac{\pi^2 n^2}{L^2}+\frac{1}{R^2}\left(\lambda+\beta\right)^2  \right]^{\frac{1}{2}}\,,
\end{equation}
corresponding to the  square integrable spinors whose components are given by Eq. (\ref{fkn1}) and Eq. (\ref{g12}) that yields now
\begin{equation}
\left(
\begin{array}{c}
g_1(z)\\
g_2(z)
\end{array}\right)=\frac{i N}{E+M}\left(
\begin{array}{cc}
- k_n\cos(k_n z)& -\frac{1}{R}(\lambda+\beta)\sin(k_n z)\\
\frac{1}{R}(\lambda+\beta)\sin(k_n z)& k_n\cos(k_n z)
\end{array}\right) \xi_{\sigma}\,.
\end{equation}
Then, according to Eq. (\ref{psi}) we can write down the form of two types of solutions corresponding to $\sigma =\pm\frac{1}{2}$. For $\sigma =\frac{1}{2}$ we obtain
\begin{equation}\label{psiUp}
U_{n,\lambda}^+(t, \phi,z)=N_{n,\lambda}^+\left(
\begin{array}{r}
\sin(k_n z)e^{i\phi(\lambda-\frac{1}{2})}\\
0\hspace*{20mm}\\
\frac{-ik_n}{E_{n,\lambda}+M}\cos(k_n z)e^{i\phi(\lambda-\frac{1}{2})}\\
\frac{i(\lambda+\beta)}{R(E_{n,\lambda}+M)}\sin(k_n z)e^{i\phi(\lambda+\frac{1}{2})}
\end{array}\right) e^{-iE_{n,\lambda}t}\,,
\end{equation}
and similarly for $\sigma=-\frac{1}{2}$,
\begin{equation}\label{psiUm}
U_{n,\lambda}^-(t, \phi,z)=N_{n,\lambda}^-\left(
\begin{array}{r}
0\hspace*{20mm}\\
\sin(k_n z)e^{i\phi(\lambda+\frac{1}{2})}\\
\frac{-i(\lambda+\beta)}{R(E_{n,\lambda}+M)}\sin(k_n z)e^{i\phi(\lambda-\frac{1}{2})}\\
\frac{ik_n}{E_{n,\lambda}+M}\cos(k_n z)e^{i\phi(\lambda+\frac{1}{2})}
\end{array}\right) e^{-iE_{n,\lambda}t}\,.
\end{equation}
After a little calculation we find that by fixing the value of the normalization constants as
\begin{equation}
N_{n,\lambda}^{\pm}=\frac{1}{\sqrt{\pi RL}}\sqrt{\frac{E_{n,\lambda}+M}{2E_{n,\lambda}}}\,,
\end{equation}
we obtain the desired orthogonality relations
\begin{equation}
\langle U^{\pm}_{n,\lambda}, U^{\pm}_{n',\lambda'}\rangle=\delta_{n,n'}\delta_{\lambda,\lambda'}\,,\quad \langle U^{\pm}_{n,\lambda}, U^{\mp}_{n',\lambda'}\rangle=0\,.
\end{equation}
 The conclusion is that the above fundamental solutions are eigenspinors of the set of commuting operators $\{H,K,J_3, P_3^2\}$.

Let us consider the fermions in the states $(n,\lambda)$ given by the normalized linear combinations  
\begin{equation}\label{psicc}
\psi_{n,\lambda}=c_+ U_{n,\lambda}^+ +c_-U_{n,\lambda}^-\,, \quad |c_+|^2+|c_-|^2=1\,,
\end{equation}
which satisfy $\langle \psi_{n,\lambda}, \psi_{n',\lambda'}\rangle=\delta_{n,n'}\delta_{\lambda,\lambda'} $. The constants $c_{\pm}$ give the polarization degree defined as in the previous case, 
\begin{equation}\label{pol}
{\cal P}=\langle \psi_{n,\lambda}, K \psi_{n,\lambda}\rangle=\lambda (|c_+|^2-|c_-|^2)\,.
\end{equation}
Obviously,  the fermions are unpolarized when $|c_+|=|c_-|=\frac{1}{\sqrt{2}}$.

With these ingredients we can calculate the quantities
\begin{eqnarray}
\overline{U}^{\pm}_{n,\lambda}\gamma^{\phi}{U}^{\pm}_{n,\lambda}&=&\frac{\lambda+\beta}{2\pi R^3 L E_{n,\lambda}}\sin^2 k_n z\,,\\ \overline{U}^{\pm}_{n,\lambda}\gamma^{\phi}{U}^{\mp}_{n,\lambda}&=&\frac{2n}{RL^2E_{n,\lambda}}\sin k_n z\cos k_n z\,.\label{usles}
\end{eqnarray}
that help us to derive the definitive form of the circular currents in the states $(n,\lambda)$ that read
\begin{equation}\label{Inm}
I^c_{n,\lambda}=R\int_{0}^{L} dz\, \overline{\psi}_{n,\lambda}\gamma^{\phi}\psi_{n,\lambda}=\frac{\lambda+\beta}{2\pi R^2 E_{n,\lambda}}=\frac{1}{2\pi}\frac{\partial E_{n,\lambda}}{\partial\beta}\,,
\end{equation}
since the integral over $z$ vanishes the mixed terms (\ref{usles}) while the constants $c_{\pm}$ satisfy Eq. (\ref{psicc}). It is remarkable that this current is independent on polarization and  has a similar form and relation with the energy as in the case of the AB rings  \cite{CBC}. The difference is that now the energy (\ref{Enm}) depends on two quantum numbers, $n$ and $\lambda$, as well as on the length $L$ of the AB cylinder. 

The longitudinal current vanishes since we used boundary conditions that guarantee that
\begin{equation}
\overline{U}^{\pm}_{n,\lambda}\gamma^{3}{U}^{\pm}_{n,\lambda}=\overline{U}^{\pm}_{n,\lambda}\gamma^{3}{U}^{\mp}_{n,\lambda}=0\,.
\end{equation}
In other words, our boundary conditions are of the MIT type vanishing the currents but without canceling  all the components of the Dirac spinors on boundaries.  

\subsection{Persistent currents on finite AB cylinders}

The properties of the currents (\ref{Inm}) can be better understood by introducing the appropriate dimensionless parameters 
\begin{equation}
\mu=MR\,,\quad \nu=\frac{\pi R}{L}\,, 
\end{equation} 
that allow us to write
\begin{equation}\label{Ichi}
I^c_{n,\lambda}=\frac{1}{2\pi R}\chi_{\mu,\nu}(n,\lambda)\,, \quad \chi_{\mu,\nu}(n,\lambda)=\frac{\beta+\lambda}{\sqrt{\mu^2+\nu^2 n^2+(\beta+\lambda)^2}}\,,
\end{equation}
pointing out the function $\chi$ which gives the behavior of the circular currents.
This function is smooths with respect to all of its variables increasing monotonously with $\lambda$ from $-1$ to $1$ since 
\begin{equation}
\lim_{\lambda\to \pm\infty}\chi_{\mu,\nu}(n,\lambda)=\pm 1\,,
\end{equation} 
and vanishing for $n\to \infty$. This means that, as in previous case, for increasing total angular momenta, the circular current tends to the asymptotic saturation values $\pm (2\pi R)^{-1}$ just  as it happens with the partial currents in AB rings \cite{CBC}.

Now we can use these properties for estimating the persistent current at $T=0$ in semiconductor AB cylinders where the electron discrete energy levels $E_{n,\lambda}$  are given by Eq (\ref{Enm}).  According to the Fermi-Dirac statistics, at $T=0$   
the electrons occupy  all the states $(n,\lambda)$  which satisfy the condition
\begin{equation}\label{Cond0}
E_{n,\lambda}\le E_F+M
\end{equation}
where $E_F\ll M$ is the (non-relativistic) energy of the Fermi level. Therefore, the total number of electrons $N_e$ and the persistent current $I$ can be calculated as
\begin{eqnarray}
N_e&=&\sum_{n,\lambda; E_{n,\lambda}\le E_F+M} 1=\sum_{n,\lambda>0; E_{n,\lambda}\le E_F+M} 2\,,\label{Ne}\\
I&=&\sum_{n,\lambda; E_{n,\lambda}\le E_F+M}I^c_{n,\lambda}=\sum_{n,\lambda>0; E_{n,\lambda}\le E_F+M} (I^c_{n,\lambda}+I^c_{n,-\lambda})\,.
\end{eqnarray}
In practice the flux parameter $\beta$ remains very small (less than $10^{-8}$) such that we can neglect the terms of the order $O(\beta^2)$ of the Taylor expansions of our functions (\ref{Ichi}).  Thus we can write 
\begin{eqnarray}
2\pi R(I^c_{n,\lambda}+I^c_{n,-\lambda})&=&\chi_{\mu,\nu}(n,\lambda)+\chi_{\mu,\nu}(n,-\lambda)\nonumber\\
&=&2j_{\mu,\nu}(n,\lambda) \beta +O(\beta^3)\,,
\end{eqnarray}
where
\begin{equation}
j_{\mu,\nu}(n,\lambda)=\frac{\mu^2+\nu^2 n^2}{(\mu^2+\nu^2 n^2+\lambda^2)^{\frac{3}{2}}}\,,
\end{equation}
obtaining thus the expression of the relativistic persistent currents,  
\begin{equation}\label{Ic}
I=\frac{\beta}{\pi R}\,c(\mu,\nu) \,, \quad c(\mu,\nu)=\sum_{n,\lambda>0;
 E_{n,\lambda}\le E_F+M}j_{\mu,\nu}(n,\lambda)\,.
\end{equation}

The principal problem in evaluating such sums is the computation of the contributing states $(n,\lambda)$ (with $n=1,2,...$ and $\lambda=\pm\frac{1}{2},\pm\frac{3}{2},...$) which  satisfy the condition (\ref{Cond0}).   We denote first by $n_F$ the greatest value of the quantum number $n$ and by $\lambda_n$ the greatest value of $|\lambda|$ for a given $n$, assuming that the states $(n_F,\pm\frac{1}{2})$ and respectively $(n,\pm\lambda_n)$ (with $n=1,2,...,n_F) $ are very close to the Fermi level, i. e. $E_{n_F,\pm\frac{1}{2}}\simeq E_{n,\pm\lambda_n}\simeq E_F+M$. In addition, we denote by $\lambda_F=\lambda_{n=1}$ the greatest value among the quantities $\lambda_n$. Then we can rewrite Eq. (\ref{Cond0}) as
\begin{equation}\label{Cond}
\nu^2 n^2+\lambda^2\le \alpha^2\,, \quad \alpha=R\sqrt{E_F(E_F+2M)}\simeq R\sqrt{2M E_F}\,,
\end{equation}
obtaining the approximative identities
\begin{equation}\label{ident}
\nu^2 {n_F}^2+\frac{1}{4}\simeq \nu^2 n^2 + {\lambda_n}^2\simeq \nu^2+{\lambda_F}^2\simeq \alpha^2
\end{equation}
that help us to estimate the numbers $n_F$ and $\lambda_n$.  Moreover, we observe that  $\lambda_F$  must be much greater than $1$ since otherwise we cannot speak about statistics. Then  we can use the approximative formula  (\ref{Bsum}) obtaining the compact results
\begin{eqnarray}
N_e&=&\sum_{n=1}^{n_F}\sum_{\lambda=\frac{1}{2}}^{\lambda_n}2=
\sum_{n=1}^{n_F}(2\lambda_n+1)={n_F}+2\sum_{n=1}^{n_F} \lambda_n\,, \\    c(\mu,\nu)&=&\sum_{n=1}^{n_F}\sum_{\lambda=\frac{1}{2}}^{\lambda_n}j_{\mu,\nu}( n,\lambda)\simeq\frac{1}{\sqrt{\mu^2+\alpha^2}}\sum_{n=1}^{n_F} \lambda_n\,,
\end{eqnarray}
that represent a very good approximation for the systems with $\mu>200$ \cite{CBC}. It remains to calculate on computer the sum over $n$ or to consider the estimation (\ref{Bsum1}) when $n_F\gg 1$.

An interesting case is of the very short cylinders with $1\ll\nu <\alpha<2\nu$ whose quantum number $n$ can take the unique value $n=n_F=1$ in order to satisfy Eq. (\ref{Cond}) that  becomes now $\lambda^2\le \alpha^2-\nu^2=\lambda_F^2$.  Consequently, the allowed states are $(1,\pm\frac{1}{2}), (1,\pm\frac{3}{2}),...(1,\pm\lambda_F)$ which means that $N_e=2 \lambda_F+1$ and
\begin{equation}\label{Ishort}
I_{short}\simeq \frac{\beta}{\pi R}\frac{\lambda_F}{\sqrt{\mu^2+\alpha^2}}= \frac{\beta}{\pi R}\sqrt{\frac{\alpha^2-\nu^2}{\alpha^2+\mu^2}}\,. 
\end{equation}
However,  for the very short cylinders with $\nu>\alpha$ the identities (\ref{ident}) do not make sense such that we need to rebuild  the entire theory without  motion along the $z$ axis ($k=0$), retrieving thus the case of the ideal AB rings \cite{CBC} for which we must substitute $\nu=0$ and $\alpha=\lambda_F$ in Eq. (\ref{Ishort}). Finally, we note that for the non-relativistic short AB cylinders with $\alpha\ll \mu$ we recover the well-known result, 
\begin{equation}
I_{nr}\simeq\frac{\beta}{\pi R}\frac{\lambda_F}{\mu}\simeq\frac{\beta}{\pi R}\frac{N_e}{2\mu}\,,
\end{equation}
of the non-relativistic persistent current in AB rings.

\section{Concluding remarks}

We presented the relativistic theory of the Dirac fermions on AB cylinders based on the complete $(1+3)$-dimensional Dirac equation with restricted orbital degrees of freedom but without affecting the spin ones. This can be achieved by using the method we proposed recently 
for the AB rings \cite{CBC} according to which the orbital restrictions must be imposes on the  Lagrangian density giving rise to a correct self-adjoint Dirac operator.

The results presented here point out two principal features of the circular currents on the AB cylinders. The first one is the relation between the circular current and the derivative of the energy with respect to the flux parameter that is the same for AB rings or cylinders either in  the relativistic approach or in the non-relativistic one. In other words this property is universal for all the AB systems with cylindric symmetry. The second feature is specific only for the relativistic circular currents that tend to saturation in the limit of very large total angular momenta, in contrast with the non-relativistic ones that are increasing linearly to infinity. 

The saturation effect determines a specific form of the relativistic persistent current on finite AB cylinders  that can be seen as a generalization of the persistent current in AB rings. Obviously, the dependence on parameters is more complicated in the case of the finite cylinders but these models are closer to the real devices involved in experiments. On the other hand, we must specify that the closed formulas derived in section 3 are only approximations that must be used prudently and completed by numerical calculations on computers.   

We believe that only in this manner,  by using combined analytical and numerical methods, we could step the threshold to the relativistic physics of the Aharonov-Bohm systems.

\appendix

\subsection*{Appendix A: Longitudinal currents}

On the infinite AB cylinders the longitudinal currents in the state $(\lambda,a)$ are determined by the structure of the wave packet (\ref{psilam}) as   
\begin{eqnarray}
I^3_{\lambda}&=&R\int_{0}^{2\pi} d\phi\overline{\psi}_{\lambda}\gamma^{3}\psi_{\lambda}=\frac{1}{4\pi}\int_{-\infty}^{\infty}dkdk' \frac{e^{it(E_{k}-E_{k'})-iz(k-k')}}{\sqrt{E_{k}E_{k'}(E_{k}+M)(E_{k'}+M)}}\nonumber\\
&\times&\left\{\left[kE_{k'}+k'E_{k}+M(E_{k}+E_{k'})\right]\left[a_{+}^*(k)a_{+}(k')+a_{-}^*(k)a_{-}(k')\right]\right. \nonumber\\
&-&\left. \frac{i(\lambda+\beta)}{R}(E_{k}-E_{k'})\left[a_{+}^*(k)a_{-}(k')+a_{-}^*(k)a_{+}(k') \right] \right\}\,,
\end{eqnarray}
where the energies are given by Eq. (\ref{Ekm}) and the arbitrary functions $a_{\pm}$ satisfy the condition (\ref{Calpha}).

\subsection*{Appendix B: Approximating sums by integrals}

Since  $\lambda_F$  must be much greater than $1$ we can use the approximative formula  
\begin{equation}\label{Bsum}
\sum_{\lambda=\frac{1}{2}}^{\lambda_n}j_{\mu,\nu}( n,\lambda)\simeq \int_{0}^{\lambda_n}d\lambda \,j_{\mu,\nu}( n,\lambda)=\frac{\lambda_n}{\sqrt{\mu^2+\nu^2 n^2+\lambda_n^2}}\simeq\frac{\lambda_n}{\sqrt{\mu^2+\alpha^2}}\,,
\end{equation}
that reproduces the numerical results with a satisfactory accuracy for $\mu>200$ \cite{CBC}. Moreover, when $n_F >100$  we can evaluate    
\begin{eqnarray}\label{Bsum1}
\sum_{n=1}^{n_F} \lambda_n&=& \sum_{n=1}^{n_F} \sqrt{\nu^2(n_F^2-n^2)+\frac{1}{4}}\nonumber\\
&\simeq& \int_{x=0}^{n_F}dx \sqrt{\nu^2(n_F^2-x^2)+\frac{1}{4}}\simeq \frac{1}{4}n_F \left(1+\frac{\pi n_F}{\nu}\right) \,.
\end{eqnarray}

\end{document}